# The influence of yaw on the unsteady surface pressures over a two-wheeled landing-gear model[1]

W. R. Graham
*Department of Engineering, University of Cambridge, Cambridge, England, CB2 1PZ, UK*

and

A. Gatto
*Department of Mechanical and Aerospace Engineering, Brunel University London, Uxbridge, England, UB8 3PH, UK*

**Abstract**

Landing-gear noise is an increasing issue for transport aircraft. A key determinant of the phenomenon is the surface pressure field. Previous studies have described this field when the oncoming flow is perfectly aligned with the gear. In practice, there may be a cross-flow component; here its influence is investigated experimentally for a generic, two-wheel, landing-gear model. It is found that yaw angles as small as 5° cause significant changes in both overall flow topology and unsteady surface pressures. Most notably, on the outboard face of the leeward wheel, large-scale separation replaces predominantly attached flow behind a leading-edge separation bubble. The effect on unsteady surface pressures includes marked shifts in the content at frequencies in the audible range, implying that yaw is an important parameter for landing-gear noise, and that purely unyawed studies may not be fully representative of the problem.

## 1 Introduction

Landing-gear noise is now widely recognised as a significant contributor to the sound radiated by transport aircraft on approach. Increasing research effort is thus being devoted to the problem; the overview of work to date given in Ref. [1] can briefly be summarised as follows. One strand of the literature, starting with the pioneering study of Heller and Dobrzynski [2], has focussed on experimental noise measurements alone. Later instances of this approach assessed the benefits of various noise-reducing modifications [3–7]. Meanwhile, other investigations [8–10] sought to understand the noise-generation process by elucidating the local flow field. More recent programmes have generated datasets which address both aspects; notable here are the LAGOON project [11–17] and the experimental part of the Rudimentary Landing Gear (RLG) initiative [18]. Computational techniques are also now routinely applied to the landing-gear noise problem; the LAGOON dataset has been the basis for significant validation efforts [12–17], as has the RLG [19, 20]. Nonetheless, the fundamental determinant of the noise radiated by the gear — the surface pressure field ([12, 13]) — is difficult to predict reliably, and this was the motivation behind the work described in Ref. [1].

[1] Some of the results in this work were presented previously in AIAA Paper 2016-2904.

A notable omission from the current literature is information on the effect of cross-flow. (Tests with the gear yawed at 5° were included in the LAGOON programme [11], but results from these configurations have not, to the authors' knowledge, been reported.) In practical operation, continuous cross-flow from side winds is usually eliminated by performing a 'crabbed' approach. Here the aircraft is yawed away from its direction of travel so that the overall oncoming flow it experiences is parallel to its axis. However, side-winds are inevitably variable to some extent, and it is possible to envisage gusts whose time-scale is too short for the aircraft orientation to adjust, but is long enough to establish an effectively steady yaw angle in the gear flow. A preliminary analysis of hourly wind data from Heathrow airport shows that ground-level speed deviations in the range 5–10kt are extremely common. Acting normal to an aircraft flying at a typical approach speed in the region of 70m/s, these would induce 2–4° yaw. Higher up, at points on the approach path above inhabited areas, greater values would be expected.

On this basis, the two-wheeled landing-gear model of Ref. [1] was tested at 5° and 10° yaw. This paper describes the results, with a focus on the aspects that differ noticeably due to the cross-flow.

## 2 Experimental setup and apparatus

### *2.1 Landing-gear model*

A schematic of the experimental model is shown in Fig. 1. The model is an idealized quarter-scale (wheel diameter 0.36m, overall width 0.5m) representation of a two-wheeled landing gear. Its main structural support strut is surrounded by an outer shroud, which contains nine planes of pressure sensors distributed along its length. Each plane has five sensors spaced at 60° intervals, and the shroud is manually rotatable by up to ±40°. A yaw adjustment plate (range ±20°) and a six-axis load cell complete the model structure.

Pressure sensors are also installed in the shroud base support, axle shroud and wheels, as shown in Fig. 2. Some of those in the base support (0–8) are further designated 'F', 'B' or 'S', signifying position on the front, back, or side respectively. The remaining 46 transducers are distributed along the axle shroud and wheels in a single plane, with spacings chosen to maximize resolution, subject to practical constraints and the degree of expected spatial variation. Each is embedded in a 4mm diameter internal cavity a short distance (typically less than 2mm) behind a 0.8mm surface orifice. The typical distance from model surface to transducer diaphragm is 3.5mm, ensuring negligible resonance and lag effects.

Full-surface mapping capability is achieved via a 1:1200 DC motor/gearbox assembly connected to the axle with a drive pulley. This allows the wheels and axle shroud to be rotated through 360°, with position monitored by a rotary encoder (Baumer BTIV 24S). A software-based integrated PID controller is used to set and maintain wheel angle during testing.

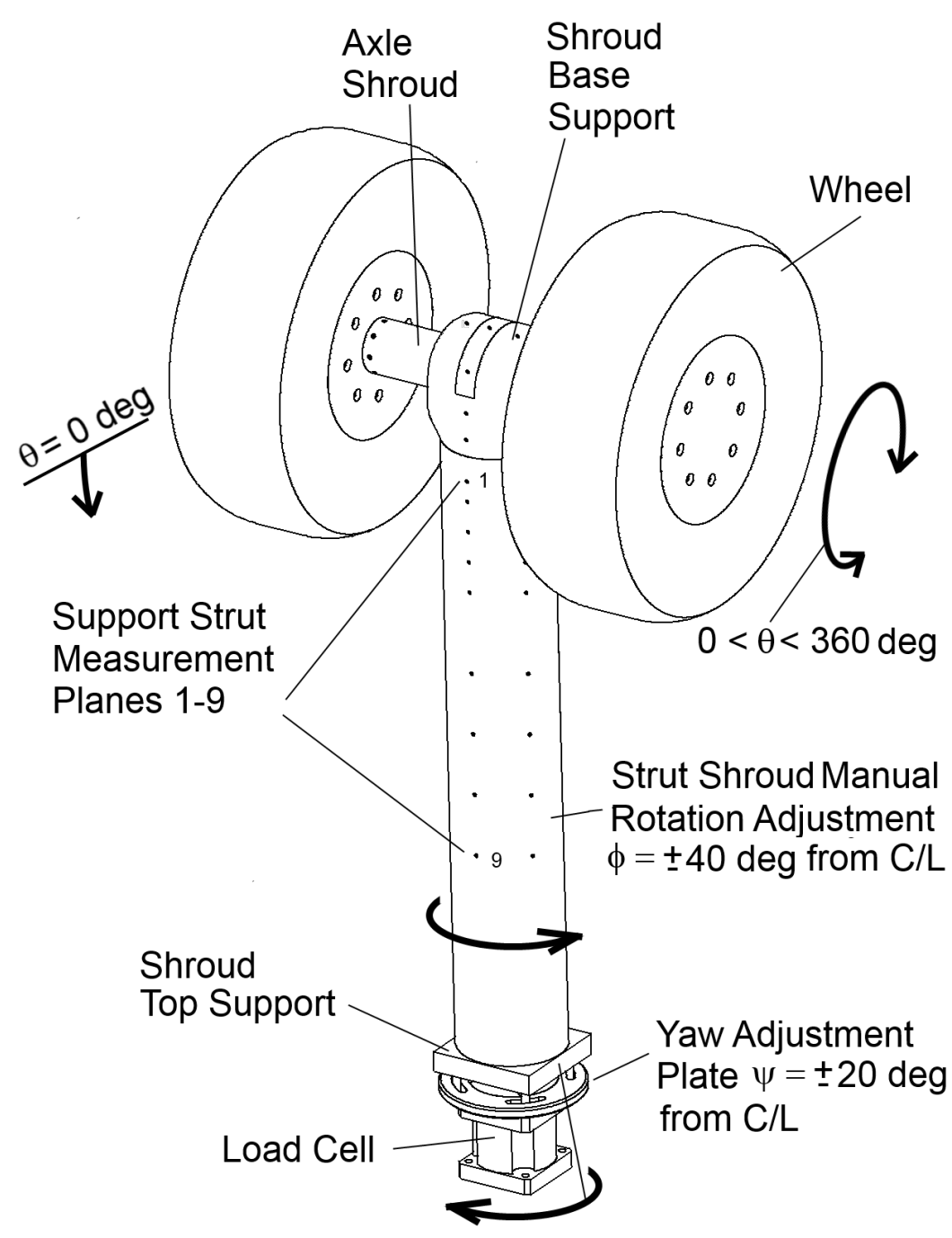

Fig. 1 Schematic of the wind-tunnel model.

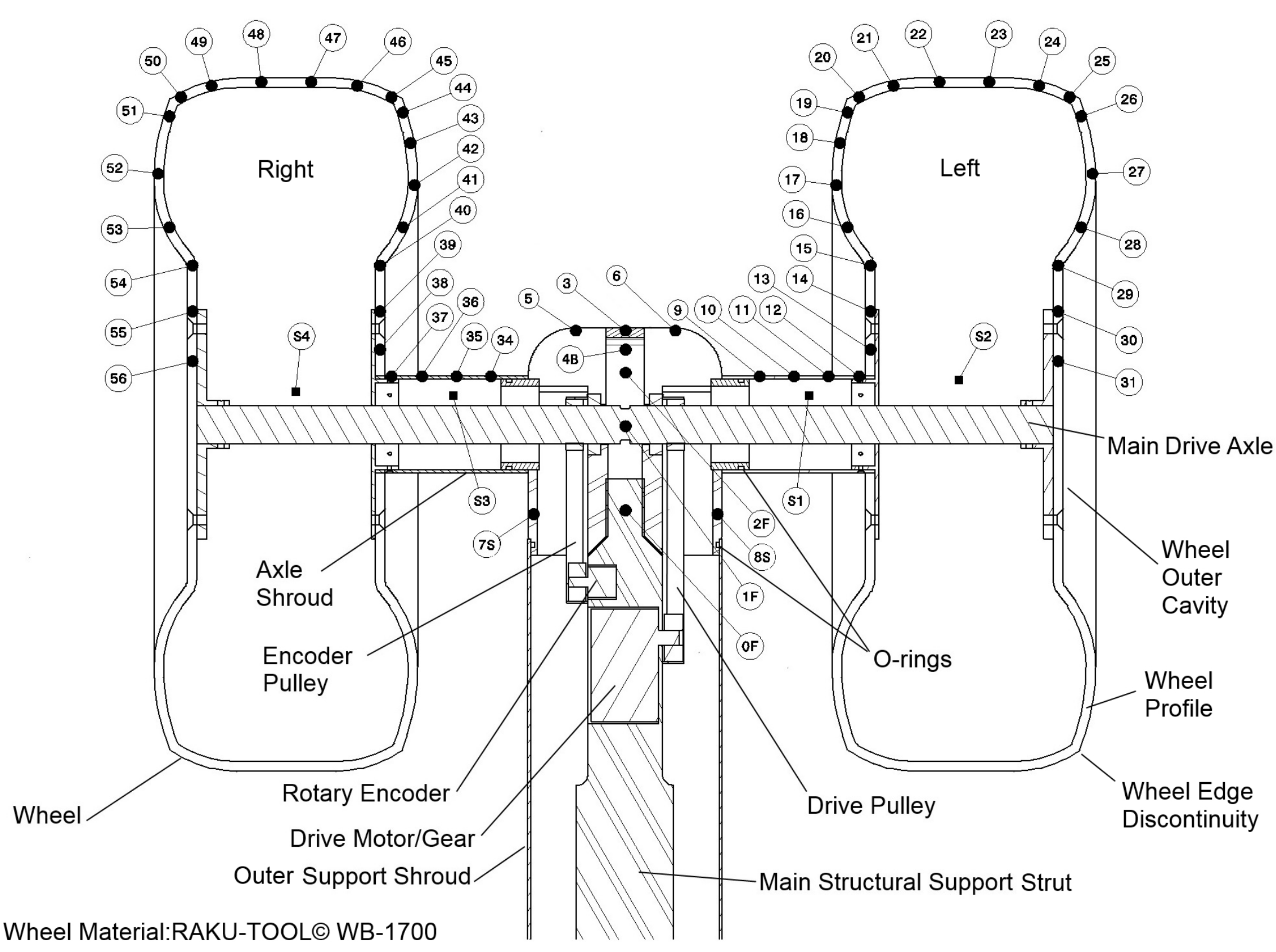

Fig. 2 Internal structure and pressure-sensor placement for landing-gear test model (front view).

To facilitate the manual rotation adjustment of the strut shroud, embedded O-rings are installed in the top and base supports (the latter near sensors 7S and 8S, as indicated on Fig. 2). Similar O-rings are also used for the axle shrouds (near transducers 9 and 34). These joints ensure an airtight seal, thereby providing a constant and uniform static pressure within the model under test conditions. This is a crucial design feature, as the (differential) transducer measurements are referenced to the internal pressure. Several additional sensors, distributed throughout the internal cavities (cf., for example, S1–S4 in Fig. 2) are available to monitor this pressure. These transducers were referenced, via pneumatic tubes, to atmospheric conditions outside the wind tunnel. Measurements during operation confirmed uniformity, with no significant difference from the ambient value. Offset corrections were therefore not applied.

### *2.2 Pressure transducers and calibration*

The pressure transducers (Honeywell type CPC03GFH) are active strain-gauge/diaphragm sensors with a cylindrical measurement port and connections for PCB mounting. A custom-designed miniature PCB is integrated onto each transducer to provide signal amplification and conditioning electronics. The resulting devices have: (1) small overall dimensions (20mm x 10mm x 7mm), allowing access to areas within the model where space is limited; (2) low capital cost compared to the industry-standard Kulite© sensors used in similar studies [11,14,18]; and (3) relatively flat response characteristics over the frequency range of interest, including the DC limit.

Each transducer was calibrated in a purpose-built rig, consisting of an evacuated chamber 25mm in length and diameter, with the facility to mount the transducer opposite a Bruel & Kjaer, laboratory-standard, 4180 reference microphone. To obtain the mean calibration, a manual pressurization port on the back of the chamber was used, with a Digitron 2081P pressure meter fitted in place of the microphone. From these mean-pressure calibration tests, typical maximum deviations within a 95% confidence interval were found to be less than ±2% of full-scale output. To assess the dynamic performance of each sensor, a generic 50mm-diameter loudspeaker was mounted directly to the front of the chamber and driven with white noise. The resulting transducer and microphone signals were then used to obtain the transducer response over the 0–6kHz frequency range.

The result of a typical calibration is shown in Fig. 3. The transducer response is near flat up to approximately 2.5kHz. From here to 3.1kHz, a reduction in amplitude ratio is observed before a significant increase up to resonance at 4.25kHz (±0.25kHz for the transducer ensemble). While it was found possible to correct for this phenomenon with reasonable accuracy up to 5kHz, an upper frequency limit of 3.5kHz was chosen to minimize complexity and error. (The variations in transducer-response amplitude are within ±1.8dB at 3.5kHz.)

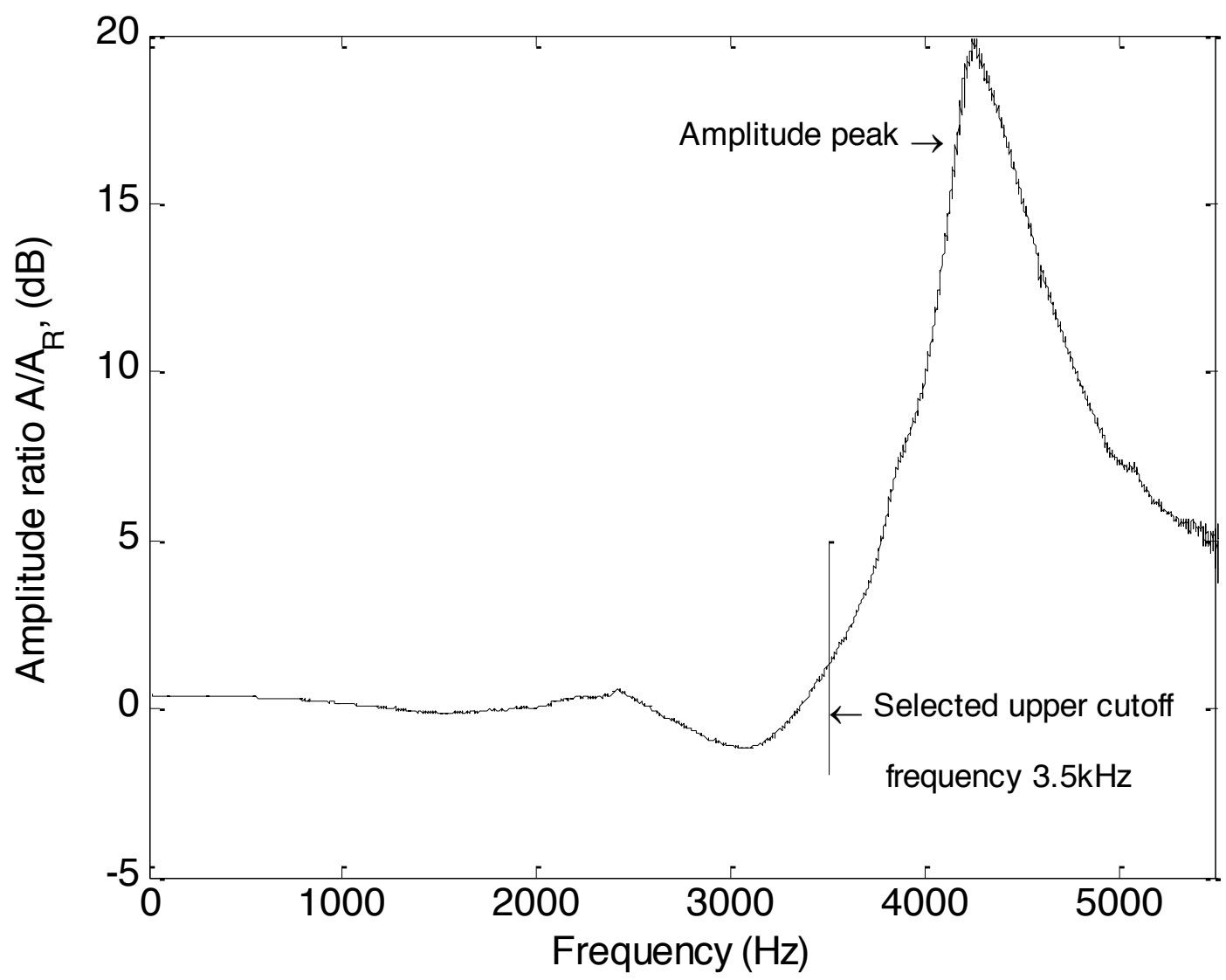

Fig. 3 Typical pressure-transducer frequency response.

### *2.3 Wind-tunnel test environment and procedure*

The model was tested in the Markham wind tunnel at the University of Cambridge. This is a closed-circuit tunnel, with a maximum flow velocity of 60m/s and a closed test section of size 1.68m x1.22m (giving a model blockage, based on frontal area, of 7.5%). The empty-section turbulence level is 0.2%. The operating flow speed was 30m/s, corresponding to a Reynolds number (based on wheel diameter) of $7.4 \times 10^5$.

The model was mounted on a support under the tunnel floor, with a two-piece, flat-plate aluminum cover placed over the exposed opening to minimize aerodynamic disturbance. A nominal 5mm gap between the support strut and the cover was provided to ensure unhindered model deflection under aerodynamic loading. Each model configuration was tested in a block of three separate wind-tunnel runs, with the strut shroud rotated 20° between each. A given run encompassed a complete rotation from θ = 0° to 360° (in 20° steps), which was then reversed to unwind the internal sensor cables.

The need for artificial transition fixing in model-scale landing-gear testing remains uncertain. Although boundary-layer trips were employed on the LAGOON [11] and 'rudimentary landing gear' [18] models, this was primarily to ensure 'CFD-friendliness' for validation purposes. Indeed, given the bluff nature of a landing gear, one would expect the state of the boundary-layer to be largely determined by the separations and reattachments that the geometry imposes. Here, therefore, no trips were installed.

Measurement signals were sampled by two linked National Instruments USB-6255 data acquisition and controller boards, after passing through purpose-built 8th-order, elliptical, low-pass integrated-circuit filters. The frequency range was controlled by a single USB-6255 analog output channel generating a TTL square wave with nominal frequency one hundred times the desired upper limit (which was set to 5.5kHz). The filtered signals were sampled at 12kHz for 10s. One of the USB-6255 boards was also used to control wheel/axle rotation, via an H-bridge motor-driver integrated circuit connected to the motor/gearbox unit. Before and after every run, a zero, wind-off, data point was taken. This allowed compensation for any thermal drift of measurement zeros during the experiment, as well as identification of superfluous, non-aerodynamic, frequency components in the data collected.

## 3 Results and discussion

### *3.1 Mean pressure distributions*

Figure 4 shows the mean pressure on the landing-gear surface. The quantity plotted is the pressure coefficient, $C_p$, defined as the gauge pressure non-dimensionalised on the free-stream dynamic pressure. Angle values in the plots refer to the azimuthal coordinate θ, which varies from 0° (upstream) to 360° through wing-side (WS, at 90°), downstream (180°) and ground-side (GS, at 270°). In the first column are the zero-yaw results reported previously [1]. There the following notable features were identified: stagnation regions on the belts (A) and axles (B); strong suction on the outboard leading edges (C); broad minima and maxima associated with the swirling wake flow (D–F); variations due to the sidewall geometry (G, K); axle influence on the inboard faces (J, L). These data imply a classical three-dimensional, vortical-wake, bluff-body flow, with separation bubbles on the outboard leading edges. Similar bubbles are either absent or negligible on the inboard leading edges, because of flow outboard deflection associated with strut and axle blockage. Reattachment after the outboard-edge bubbles is implied by the subsequent pressure recovery, and has been confirmed via other measurements [1]. On the same basis, the outboard-face hub regions were identified as weakly separated. Figure 5(a) presents a schematic summary of this description.

The second and third columns in Fig. 4 show the corresponding results for, respectively, 5° and 10° yaw. (The lateral velocity component is in the negative-y direction.) There is relatively little change in the upstream belt regions, apart from a slight stagnation-point shift on the leeward wheel (A*). However, yaw has a marked effect on the outboard-leading-edge suction regions; on the windward wheel the suction peak is attenuated (cf. Figs. 4(a), (f)), and on the leeward wheel it disappears entirely (Figs. 4(a), (e)). Equally, so does the pressure recovery on the outboard face of this wheel. The clear implication is that the flow now separates at the beginning of the belt/sidewall transition, and remains separated. (The lack of a clear adverse pressure gradient before the postulated separation point does not invalidate this interpretation; it simply means the separation must be of sharp-edge form. An argument for attached flow around the edge cannot be supported in the absence of significant suction at this point.)

Two other, less marked, differences should also be noted. First, the initially small suction peak on the inboard leading edge of the windward wheel develops to become a more notable feature (C*). Also on this wheel, the suction associated with flow accelerating around the inner edge into the wake (F) is reduced (F**), while a new counterpart (F*) appears on the outer edge. These changes are consistent with the modified direction of the oncoming flow.

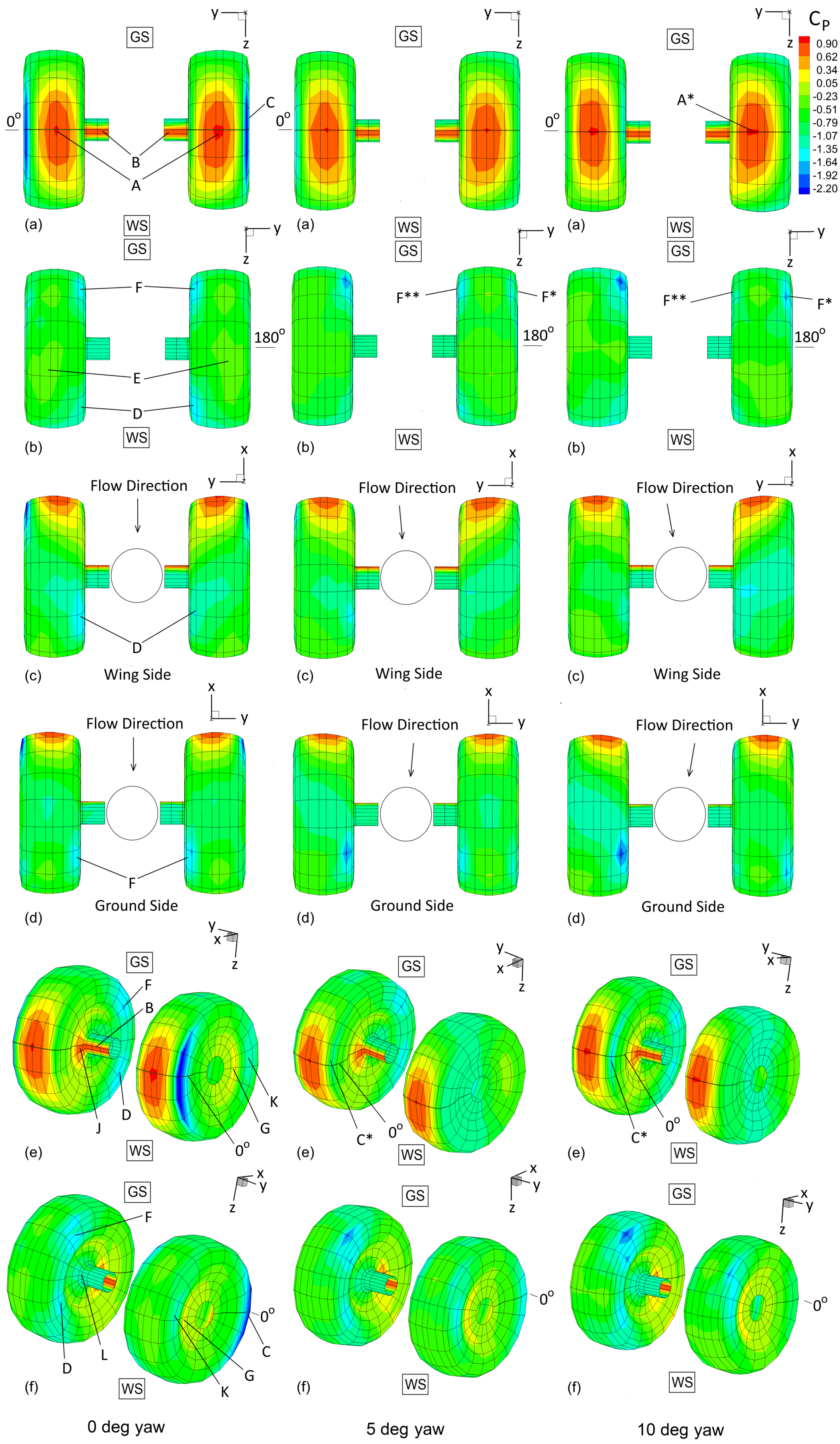

Fig. 4 Iso-contours of mean pressure coefficient over the wheel and axle assemblies.

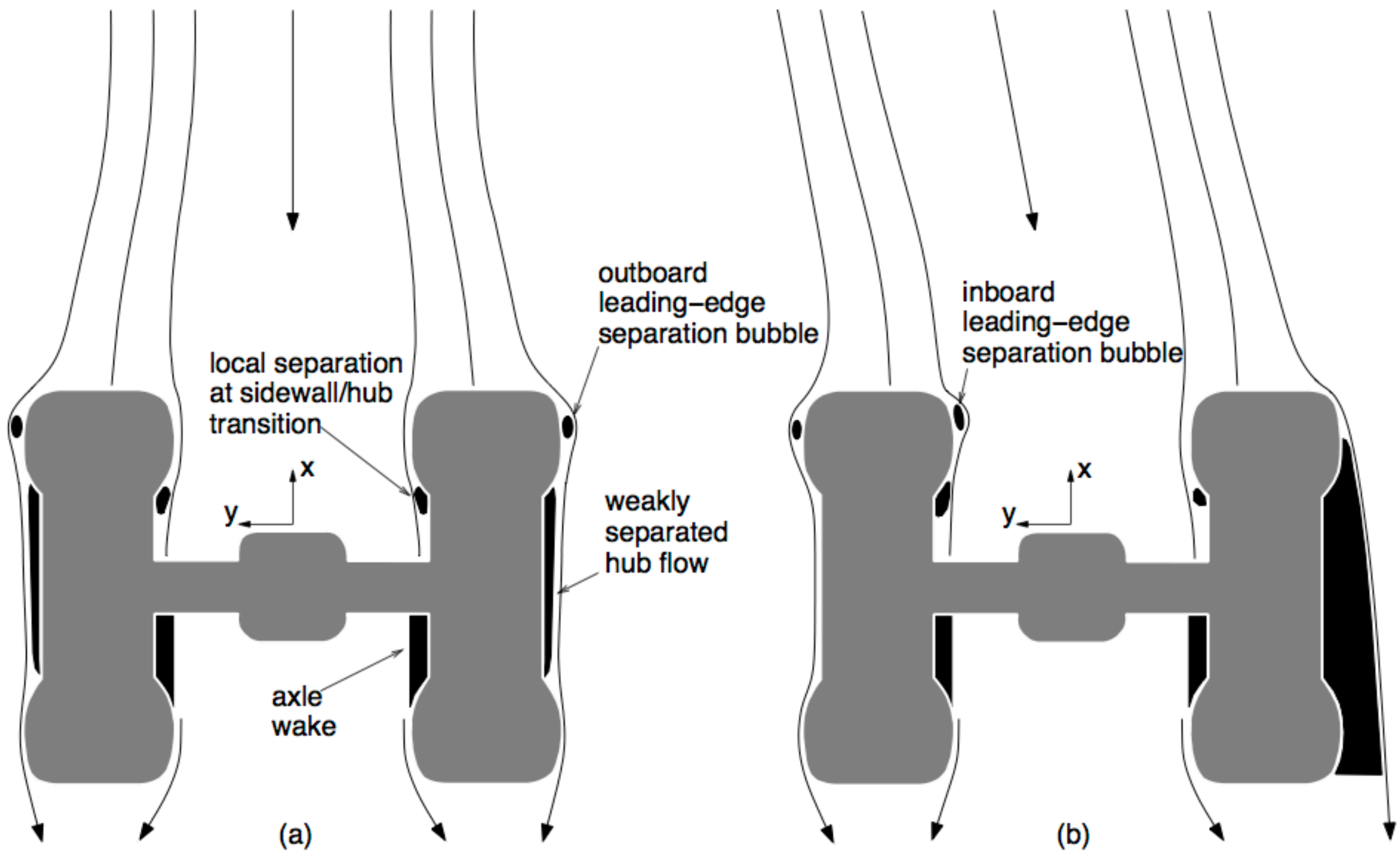

Fig. 5 Schematic of the flow in the gear mid-plane region: (a) unyawed; (b) yawed. Separated regions adjacent to the surface are shaded black. Wakes downstream of wheels and axle are not shown.

Further insight into the leading-edge and wheel-face flows is provided by Fig. 6, which shows line plots of the mean pressure along the θ = 0° position. The curves on the left (transducers 34–56) are for the windward wheel, and those on the right (transducers 9–31) for the leeward. The collapse of the outboard suction peak on the leeward wheel is immediately evident, as is its more gradual reduction on the windward wheel. This wheel's inboard-edge suction-peak growth can also be observed. More subtly, the corresponding peak on the leeward wheel reduces as the model is yawed.

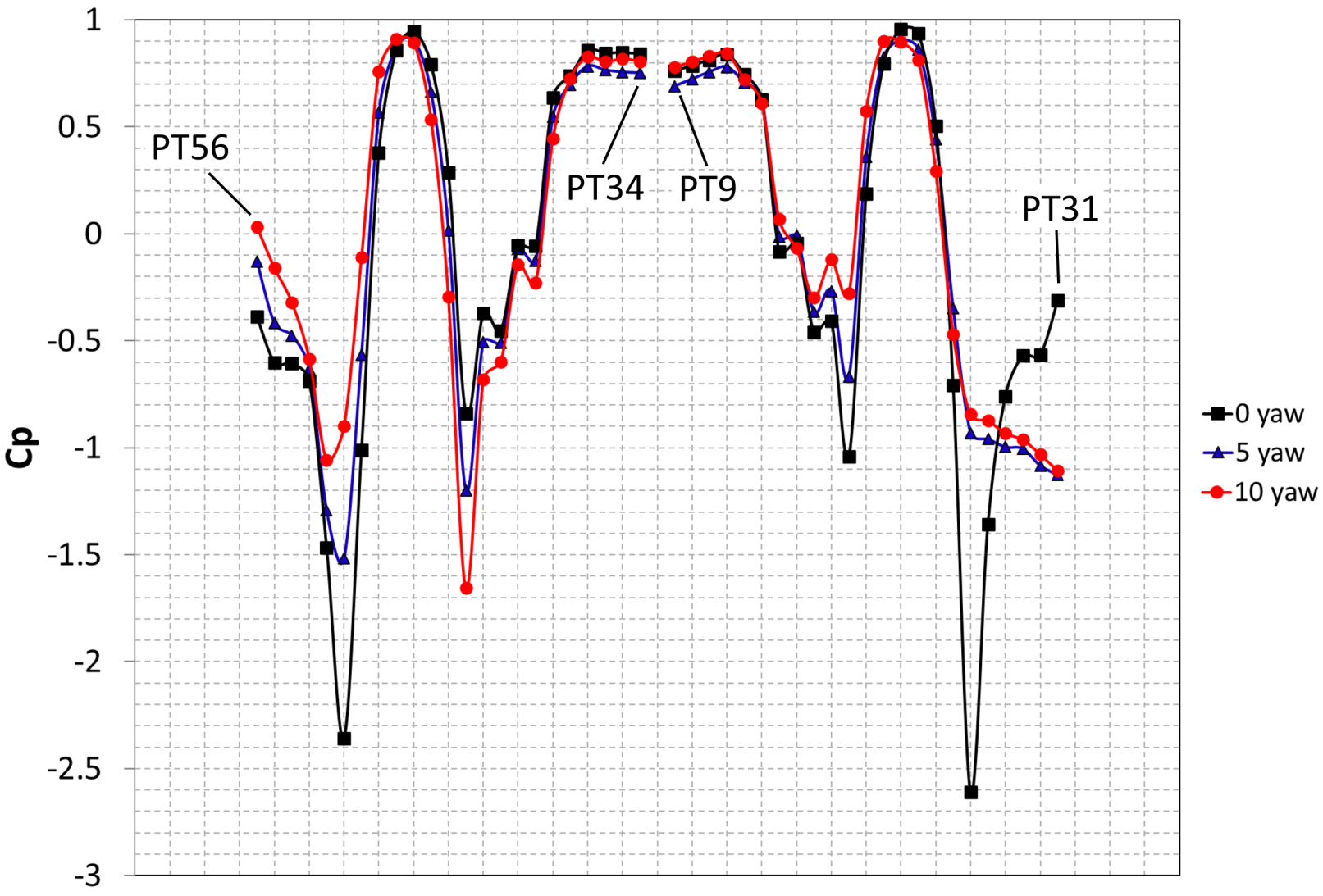

Fig. 6 The influence of yaw on the mean pressure coefficient over the wheels and axles for θ = 0°. Left side is windward, and right leeward.

Finally, Fig. 7 shows the mean-pressure line plots at θ = 220°. On the windward wheel, the increase in suction at F* is evident for transducer 50, and the reduction at F** covers transducers 42–45. For the leeward wheel, transducers 25–31 clearly show the dramatic impact of yaw on the outboard-face flow.

The yawed-flow topology deduced from these observations is shown schematically in Fig. 5(b). Note the inboard-leading-edge separation bubble on the windward wheel. This is inferred in part from the aforementioned suction peak there, but also from other evidence presented subsequently.

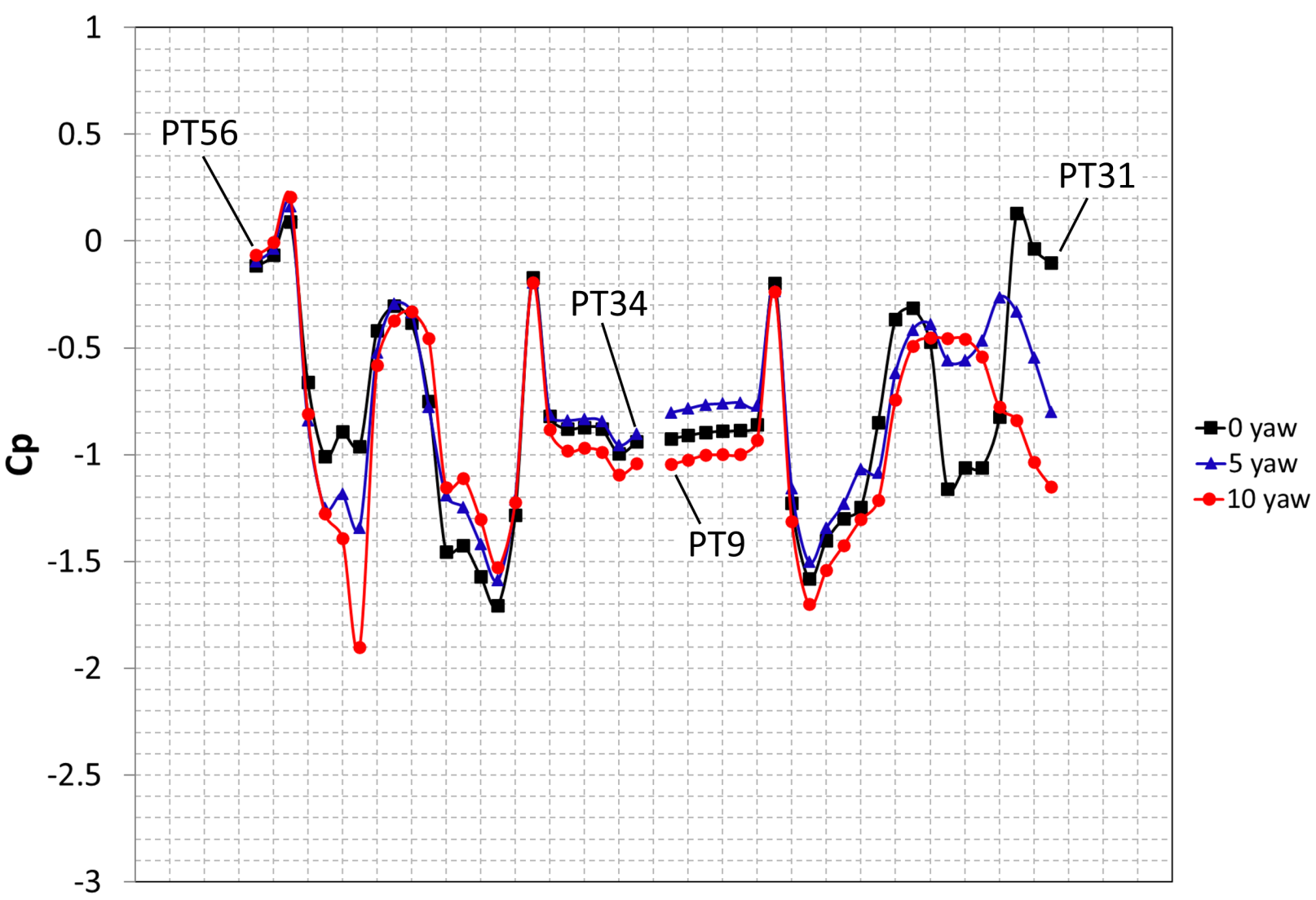

Fig. 7 The influence of yaw on the mean pressure coefficient over the wheels and axles for θ = 220°. Left side is windward, and right leeward.

### *3.2 Root-mean-square pressure distributions*

The rms-pressure results are shown as surface plots in Fig. 8. On the leeward wheel, the high levels associated with the outboard-leading-edge separation bubble (A) are eliminated by yaw (A*), suggesting that the flow, although separated, is now less energetic. Downstream, the unsteady pressures no longer drop sharply (B), instead rising gradually over the region denoted by B*. This provides further support for persistent flow separation here. Interestingly, though, the levels are higher at 5° yaw than at 10°. It seems unlikely that the separated flow is significantly less energetic at the larger angle, so this probably implies that the regions with strong unsteadiness are simply further from the wheel, and hence have less impact on the surface pressures.

Meanwhile, on the windward wheel, the continued presence of a separation bubble at the outboard leading edge is confirmed by the persistence of raised rms levels in this area (C). However, the magnitude reduces with increasing yaw, in line with the behaviour of the mean-pressure suction peak. Downstream, there is a slight reduction in unsteadiness on the outboard face (D*), consistent with elimination of the weakly separated hub flow that is present at 0° yaw (D). Such a development would be a plausible outcome of the change in oncoming flow direction.

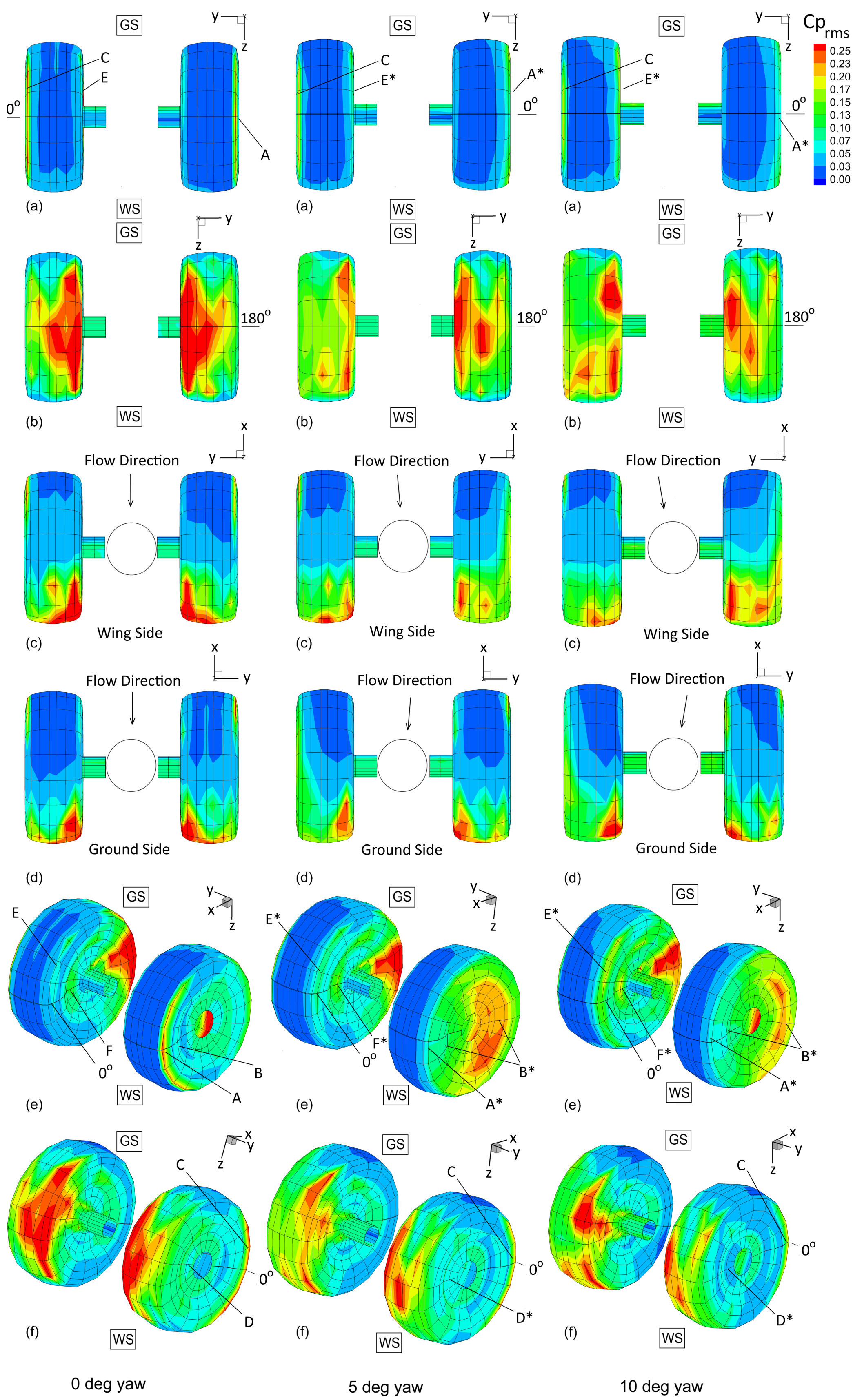

Fig. 8 Iso-contours of rms pressure coefficient over the wheel and axle assemblies.

On the inboard leading edge, the previously noted growth of a suction peak appears not initially to be matched by a comparable increase in unsteadiness; the rise (from E to E*) only becomes obvious at 10° yaw. However, closer examination of this region on the θ = 0° line (Fig. 9, transducer 43) shows that the rms pressure does in fact also increase between 0° and 5° yaw, suggesting that a separation bubble is now definitely present. The outboard-leading-edge rms, on the other hand, decreases rather little with yaw, and much less than the mean suction there (cf. Fig. 6). Also evident on this wheel are increased PT 39 levels (F*), as the flow passes from the sidewall to the hub on the inboard face. This is a region where separation might be expected, so the presence of a peak is unsurprising, but its growth with yaw (from F) is interesting because the mean pressure gradients in this region appear not to change significantly (Fig. 6). Hence the incoming flow must be more vulnerable, due to having undergone the altered leading-edge conditions.

Finally, examination of the downstream belt regions in Fig. 8 (especially on the leeward wheel) shows that yaw also affects the wheel wakes. In general, it is difficult to give a straightforward characterisation of the changes, but note that the increase in mean suction at F* (Fig. 4) corresponds to raised rms levels in this region. Similarly, but more subtly, the decrease in mean suction at F** is associated with a slight drop in rms levels.

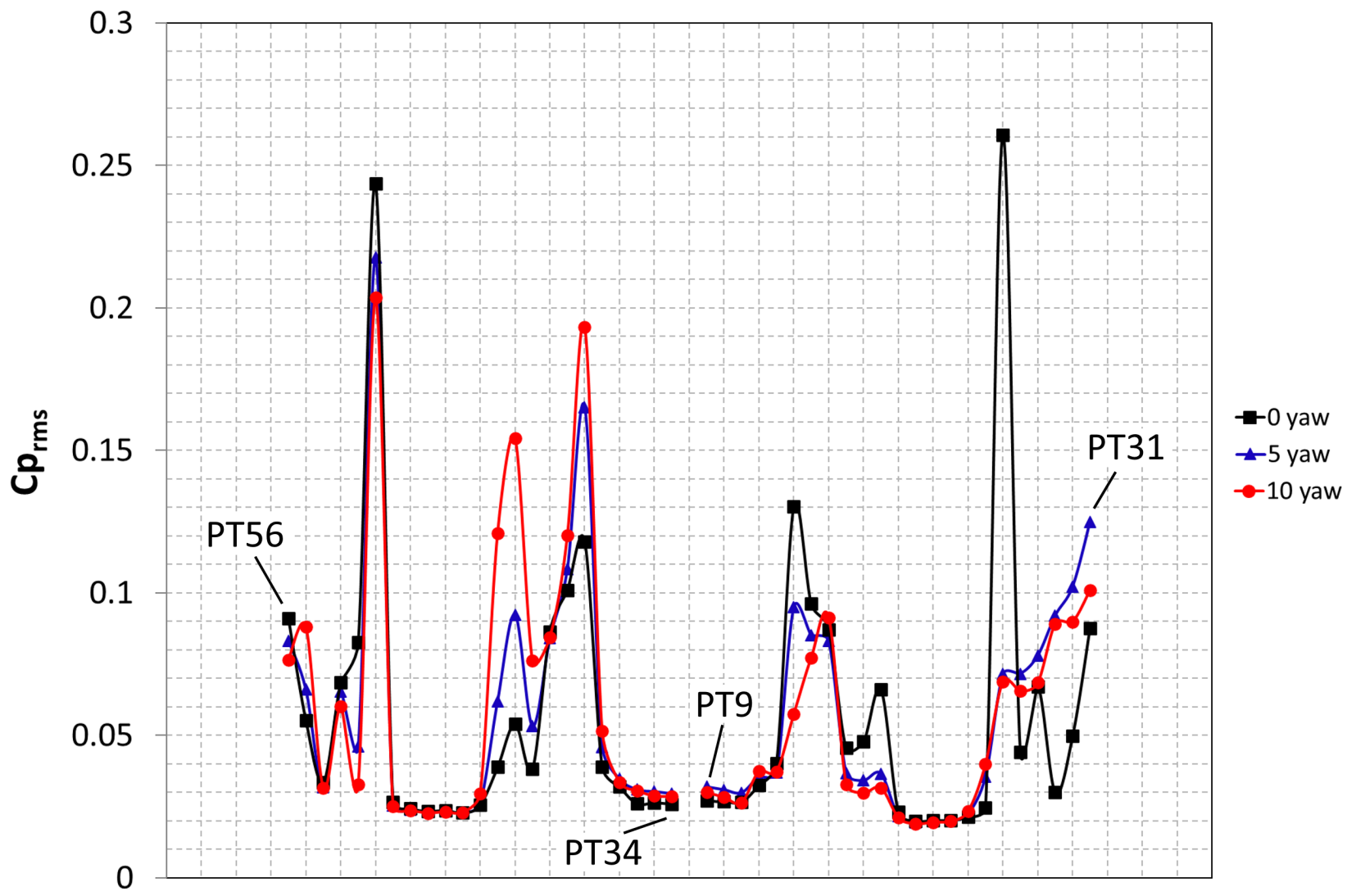

Fig. 9 The influence of yaw on rms pressure coefficient over the wheels and axles for θ = 0°. Left side is windward, and right leeward.

### *3.3 Pressure spectra*

For reasons of space, only a subset of transducer spectra is presented here. The instances chosen either show extensive variation with yaw, or illustrate a point of specific interest. The remaining, omitted, spectra are only weakly affected by yaw. To allow visual comparison of the relative importance of different frequency ranges, the results are plotted as energy per unit of log-frequency (cf. Appendix).

### *3.3.1 Inboard edges*

Spectra at the wheel inboard edges are shown in Fig. 10, which plots the results for PT 19 (leeward) and PT 44 (windward). As might have been expected from the mean- and rms-pressure results, yaw has little effect in the downstream arc (from 90° to 270°); only on close inspection are the changes associated with the region F** (on Fig. 4) visible as reduced PT 44 low-frequency levels around θ = 220°. However, notable differences are evident upstream for this transducer, and these are consistent with the postulated leading-edge flow evolution, from attached to separation-bubble. On the leeward wheel, the benign influence of cross-flow is manifested in reductions in high-frequency levels at upstream locations.

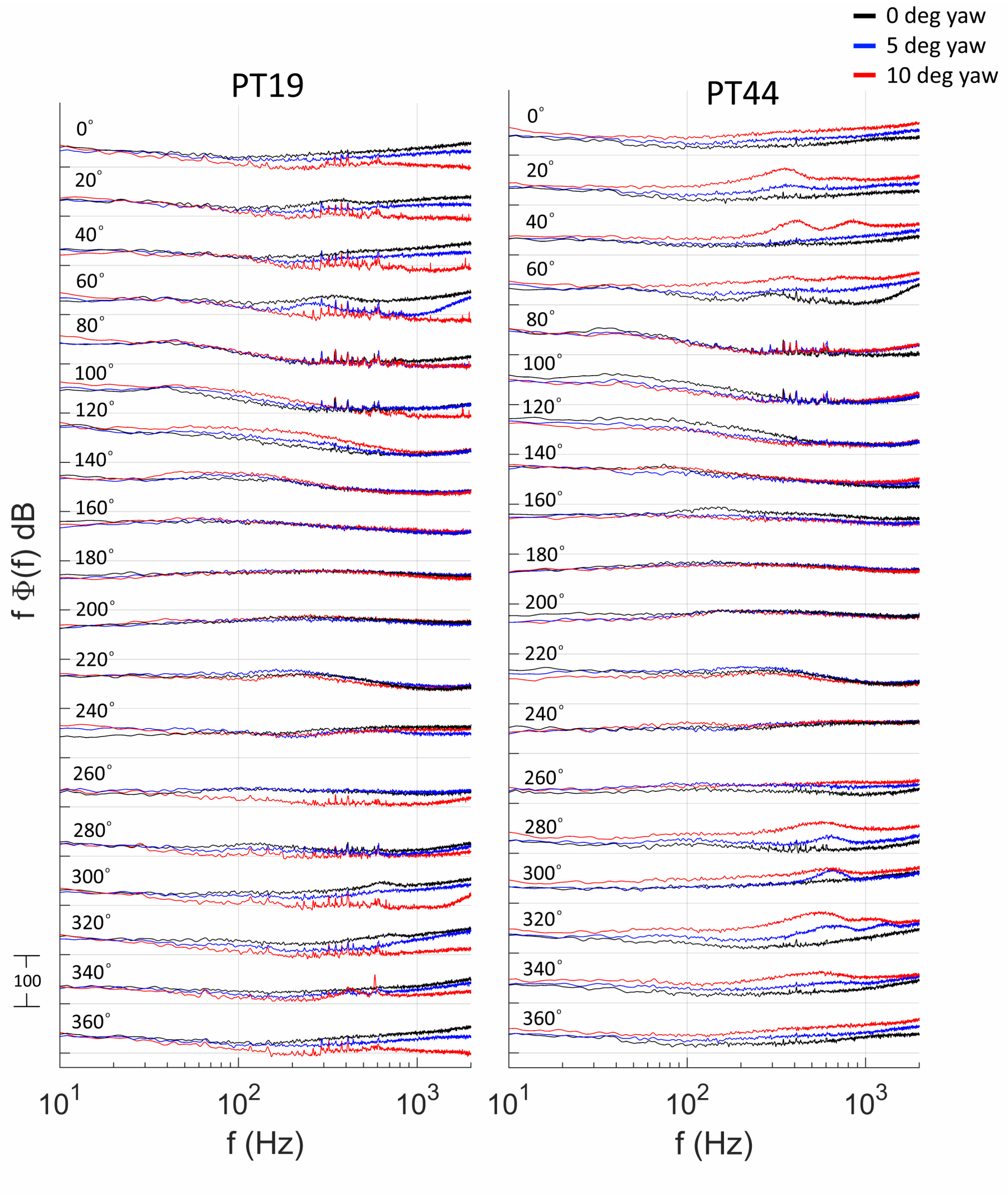

Fig. 10 Influence of yaw on the pre-multiplied pressure spectra at the wheel inboard edges.

### *3.3.2 Belt regions*

Figure 11 shows the spectra for the transducers just outboard of the belt centre-lines, PT 23 (leeward) and PT 48 (windward). Although rms-level variations in other parts of the belts are evident in Fig. 8, they typically correspond to spectrum changes at the lowest frequencies only, without other notable features (cf., for example, the data for PT 48 at θ = 200°). Transducers 23 and 48, however, are of interest because in the unyawed case they exhibit raised high-frequency levels at θ = 120° and θ = 240°. These correspond to 'inner vortex rollup attachments' [1], i.e. impingement regions for the flow swirling into the wake from between the wheels. The spectra for PT 48 at θ = 120° show a significant high-frequency drop due to yaw, which is consistent with the change in direction of the oncoming flow (tending to oppose swirling from the inner region back onto this wheel's belt). Interesting, though, is the lack of commensurate reduction at θ = 240°. This suggests that the increased mean suction and rms pressure outboard of this region corresponds to the development of an outer vortex, whose flow impingement replaces that of the previous inner vortex.

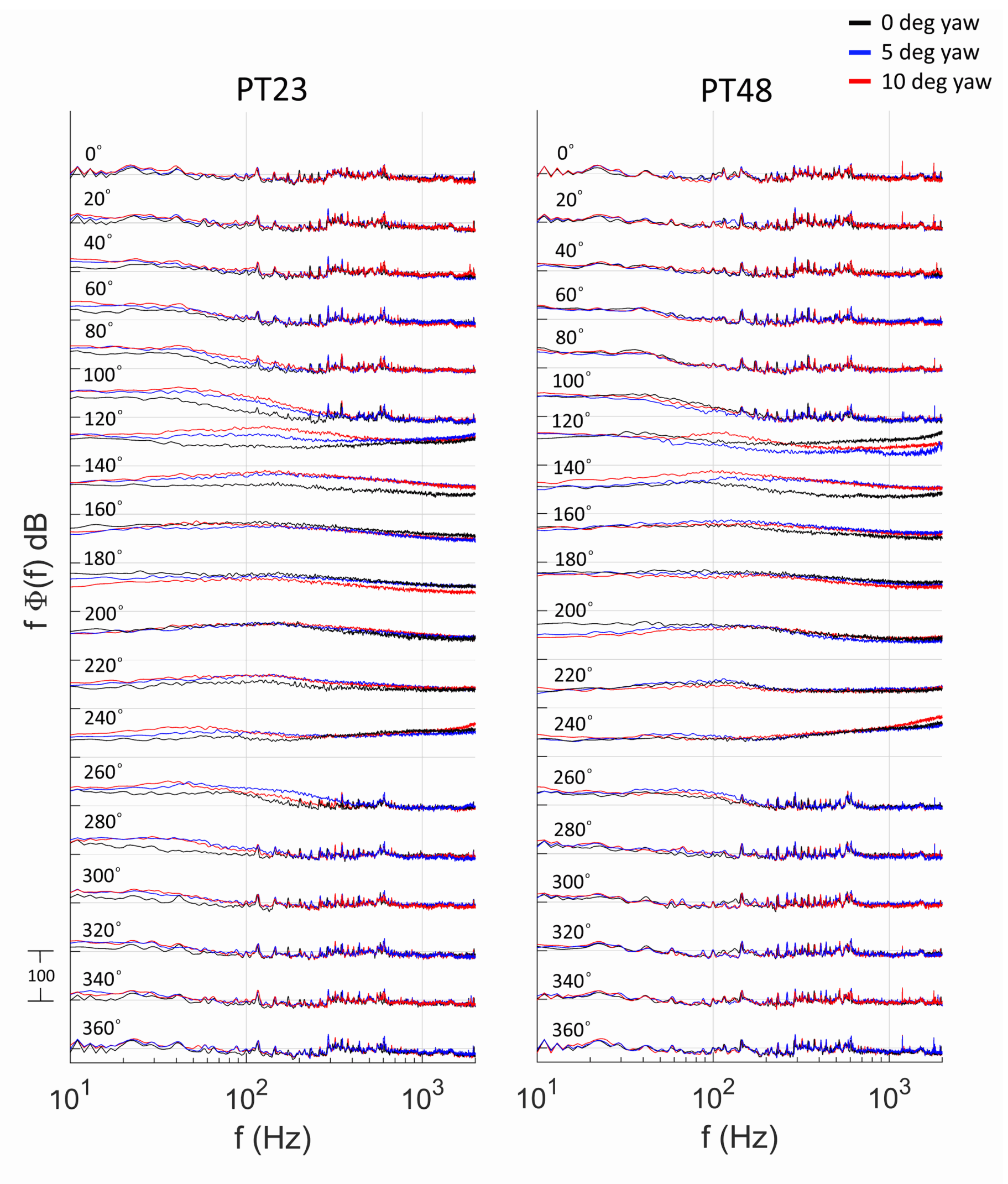

Fig. 11 Influence of yaw on the pre-multiplied pressure spectra at the belt centres.

On the leeward wheel, one would expect the inner vortices still to be present and, if anything, strengthened by yaw. Certainly, the PT 23 spectra at θ = 120° and 240° confirm their continued existence, but there is little change in high-frequency levels. There are, however, noticeable mid-frequency increases at and around these angles.

*3.3.3 Outboard edges and faces*

Figures 12–15 show the transducer spectra from the belt outboard edges to the outer hub centres. The leeward wheel is considered first. Figure 12 presents results from the belt/sidewall junction region. Transducer 25 is located at the very edge of the belt. Here all but the downstream-quadrant spectra exhibit marked increases with yaw across almost all frequencies. This is consistent with the deduction from the mean- and rms-pressure results, namely earlier flow separation in the yawed cases. In contrast, only faint, high-frequency, traces of the developing leading-edge separation bubble are visible in the unyawed spectra.

By PT 26, on the sidewall, the unyawed flow is also separated in the upstream quadrant, but its spectra remain markedly different; the separation-bubble flow has more high-frequency content, and less low, than the fully separated yawed configurations. Also evident in the latter are significantly raised levels around 90° and 270°, where the flow in the unyawed case was attached. Finally, in the downstream quadrant, all cases have high, and comparable, levels; even the unyawed flow separates at the trailing edge.

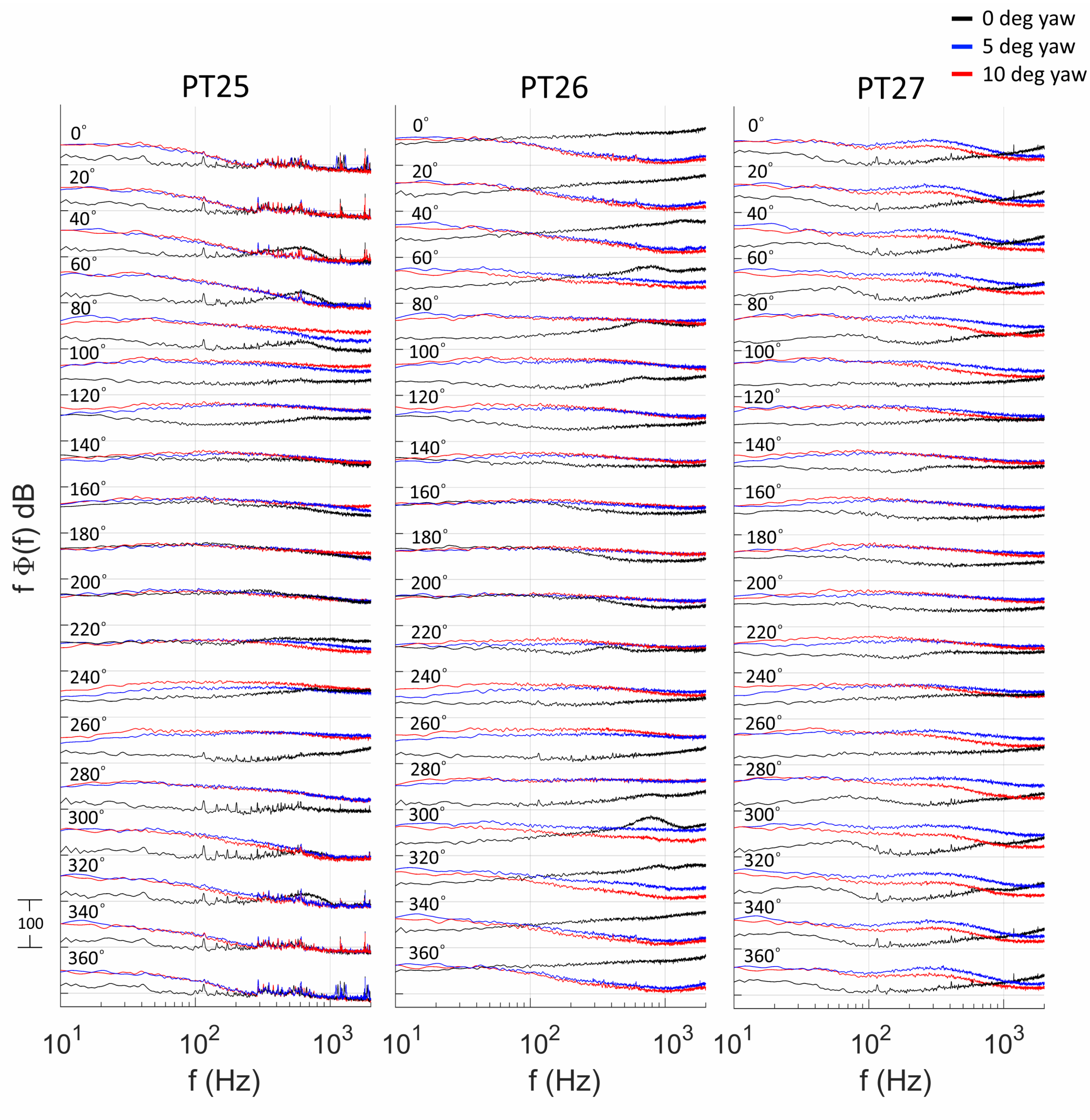

Fig. 12 Influence of yaw on the pre-multiplied pressure spectra on the leeward-wheel outboard edge.

This aspect changes markedly once the transition from belt to sidewall is complete (PT 27); now the yawed spectra are notably higher, due to the associated switch from attached to separated flow on this face. The differences persist to the upstream positions, where they are, if anything, even greater.

Transducer 28 shows similar characteristics to PT 27, but PTs 29–31, in the hub region, differ (Fig. 13). This is the area identified as weakly separated in the unyawed case, and the high-frequency levels are essentially unaffected by the departure to yawed flow. The low-frequency unsteadiness, however, is markedly higher, confirming the change in character of the separation.

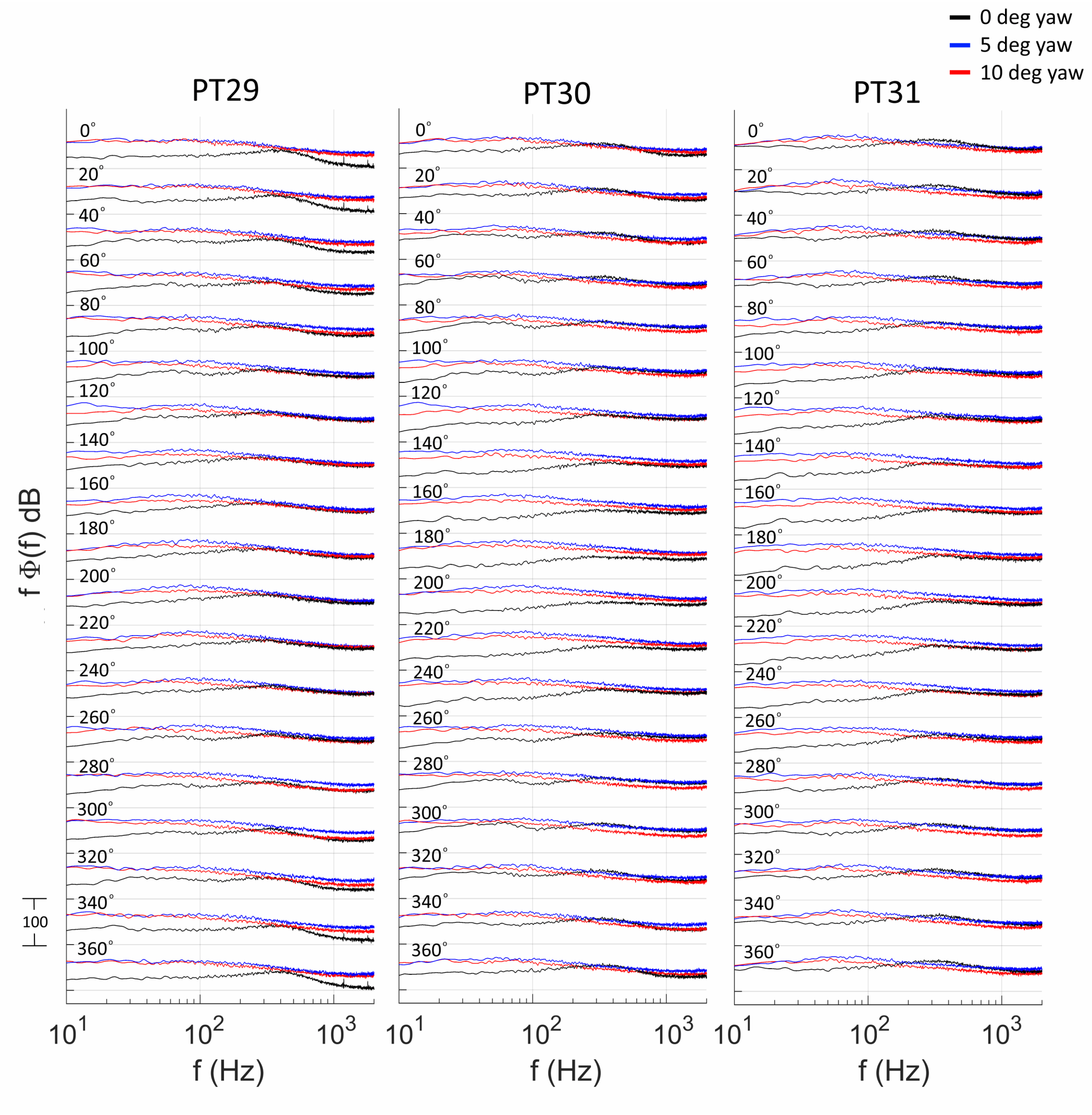

Fig. 13 Influence of yaw on the pre-multiplied pressure spectra on the leeward-wheel outboard face.

Turning to the windward wheel, Fig. 14 presents belt/sidewall-region data. The PT 50 and PT 51 spectra in the upstream arc show more clearly than the overall rms data that yaw diminishes the intensity of the leading-edge separation bubble. In the downstream arc yaw has little effect. This is true also on the sidewall (PT 52), where the spectra show similar variation with yaw. (Note, however, that high-frequency levels are now uniformly lower in the upstream locations.) However, around the sidewall/hub junction (Fig. 15, PT 53), downstream variations become evident, as would be expected if yaw eliminates the weak hub separation on this face. The extent of the region with significant variation then increases towards the hub centre, becoming approximately 60°–300° for PT 54, and covering the entire azimuth at PT 56.

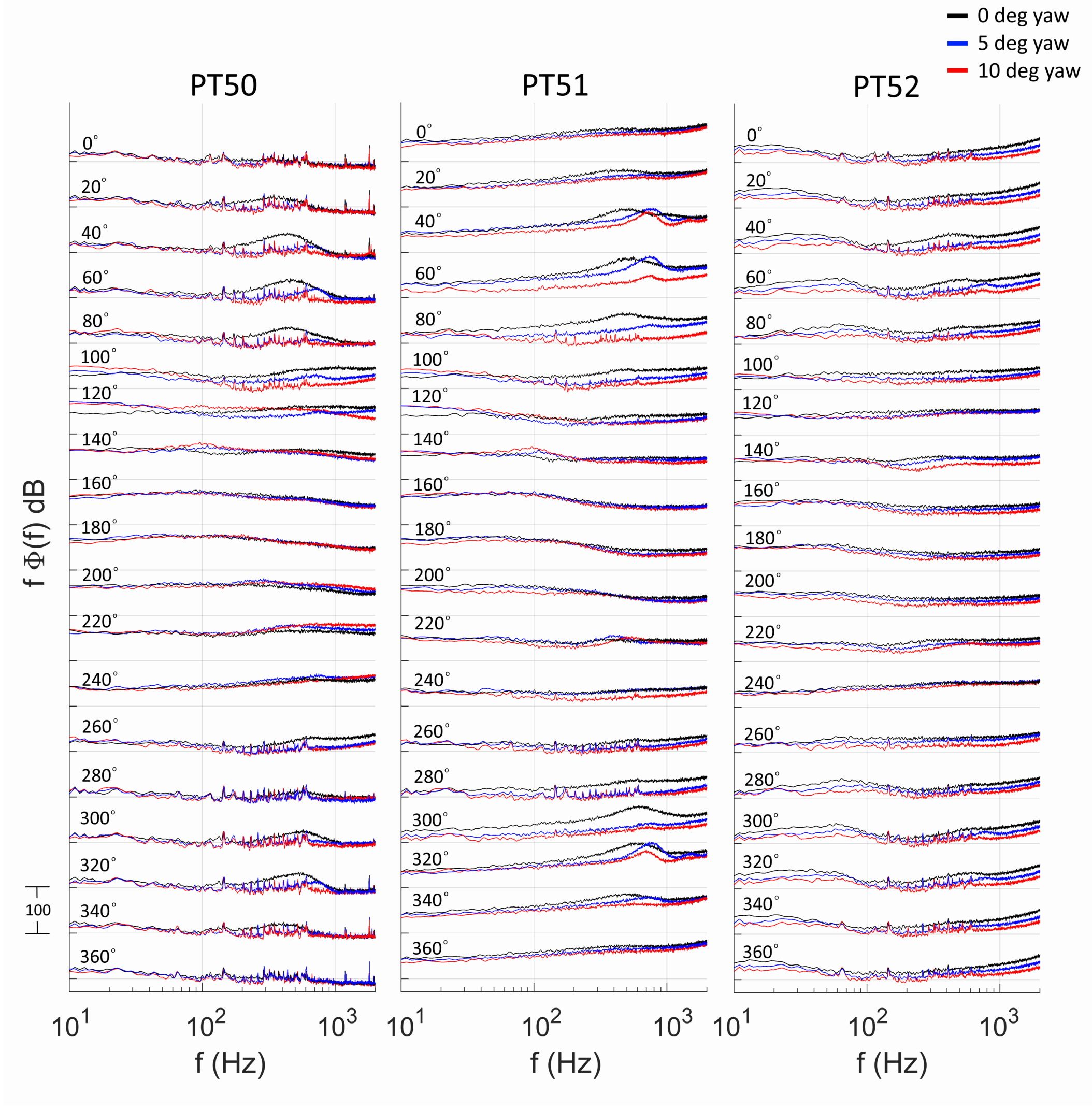

Fig. 14 Influence of yaw on the pre-multiplied pressure spectra on the windward-wheel outboard edge.

### *3.4 Practical implications*

The results presented here provide evidence of both progressive and abrupt flow-topology alterations in response to small departures from the unyawed configuration. In the former category are the changes to the wake structure downstream of the wheels, the evolution of a separation bubble on the inboard leading edge of the windward wheel, and the corresponding attenuation of this wheel's outboard-leading-edge separation bubble. In the latter are the transition from attached to separated flow on the outboard face of the leeward wheel, and the associated disappearance of its outboard-leading-edge separation bubble.

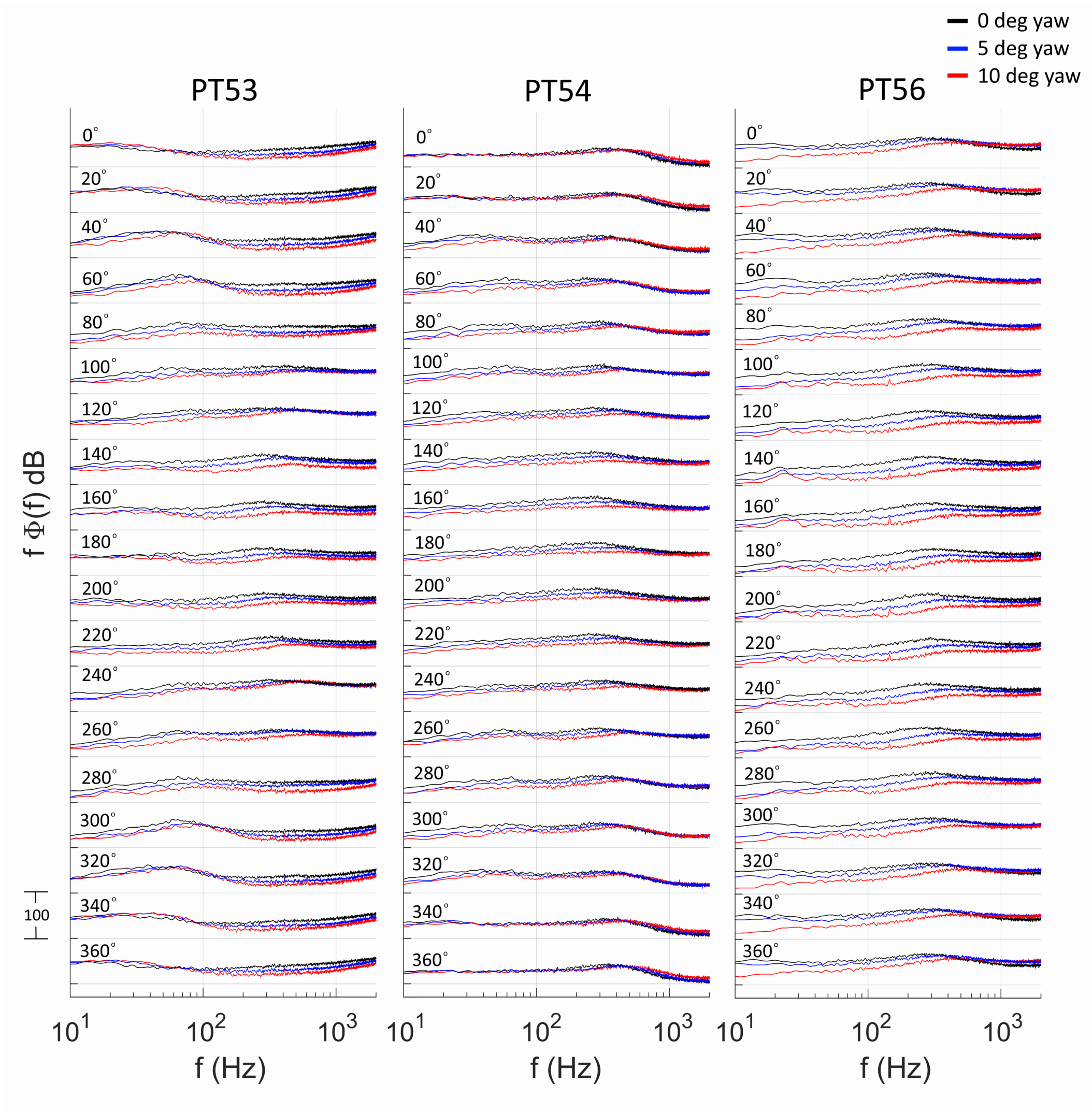

Fig. 15 Influence of yaw on the pre-multiplied pressure spectra on the windward-wheel outboard face.

All these alterations are likely to affect noise radiation, as they involve changes in the degree, extent, and characteristics of the unsteady surface pressures. The nature of the impact is uncertain, and may be counter-intuitive. For example, the yaw-induced separation on the outboard face of the leeward wheel leads to a marked increase in rms pressure, but much of this is associated with frequencies below 200Hz. On the basis of a straightforward Strouhal scaling (frequency proportional to flow speed divided by size), the corresponding full-scale components would manifest at less than 100Hz, where the A-weighting curve [21] shows a marked reduction in the ear's sensitivity to sound. On the other hand, the simultaneous elimination of the leading-edge separation bubble on this face affects rms levels over a much smaller region, but the reduction is associated with higher, more audible, frequencies. Hence, although yaw has increased the overall amount of separated flow, it might result in reduced noise levels. Finally, it must always be borne in mind that the link between surface pressures and far-field sound is complex. Even for an isolated gear, spatial correlations play an important rôle. In the real, installed, configuration, reflection and diffraction by other airframe components are also potentially important.

At this stage, it is worth recalling a key conclusion of the original study [1]: future work should aim to ascertain and reproduce boundary-layer states at full scale, because this detailed information may have a significant bearing on the noise field. To this point can now be added the probable influence of boundary-layer state on the yawed-flow topology, given the sensitivity observed here. More subtle Reynolds-number effects might also arise. For example, even if (as was claimed possible in Sec. 2) the belt stagnation regions at full scale follow this experiment in being laminar, it seems unlikely that the angle at which the flow switches away from the unyawed topology will be independent of Reynolds number.

## 4 Conclusions

This paper has described the effect of yaw on wind-tunnel measurements of landing-gear surface pressures. The landing-gear model is a simplified representation at quarter-scale; the Reynolds number is approximately one-tenth of full scale.

The key finding is that small yaw angles lead to large changes in flow topology. The most notable are on the outboard face of the leeward wheel. Without yaw, the flow is largely attached downstream of a leading-edge separation bubble. For both yaw angles tested — 5° and 10° — the flow separates irrevocably at the leading edge. The corresponding region on the windward wheel is less obviously affected, but the degree of leading-edge suction reduces and a weak hub separation is eliminated. There are also modifications to the flow between the wheels, and to the wake. Of the former, the most prominent is the development of a leading-edge separation bubble on the inboard face of the windward wheel.

Associated with these changes are significant alterations to the mean and unsteady pressure fields. Furthermore, spectral analysis shows that the latter are manifested at frequencies that would be in the audible range at full scale. It is thus likely that the radiated noise would be similarly affected.

The extent to which quasi-steady yaw at the level investigated here is encountered in day-to-day operation remains uncertain. However, Heathrow wind data for 2016 suggest that 5° would not be an exceptional occurrence. Hence it appears that future work should always test, and document, the influence of yaw.

## Acknowledgements

This research did not receive any specific grant from funding agencies in the public, commercial, or not-for-profit sectors. The implicit financial support provided by the authors' respective institutions is gratefully acknowledged. The authors also thank Duncan Ball of the UK Met Office for supplying Heathrow wind data.

## Declaration of Conflicting Interests

The Authors declare that there is no conflict of interest.

## Appendix

On spectrum plots with logarithmic frequency scales, it is difficult to gauge the relative importance of contributions from different ranges to the overall signal power. The problem can be eliminated by pre-multiplying the spectrum by frequency. This weights the spectrum level so that integration 'by eye' (as intuitively performed when viewing linear-scale plots) is again possible. The demonstration follows from writing the link between the spectrum, $\Phi$, and the mean-square pressure, $\overline{p^2}$, in the form

$$\overline{p^2} = \int f\Phi(f)\frac{\mathrm{d}f}{f}, \quad (1)$$

where $f$ is the frequency. Now $\mathrm{d}f/f = \mathrm{d}(\log f)$. Hence, for two 'bins' with equal increments on the logarithmic frequency scale, the respective values of $f\Phi(f)$ can be compared directly to assess their contributions to $\overline{p^2}$.